\documentclass[%
reprint,
nofootinbib,
amsmath,amssymb,
aps,
]{revtex4-1}
\usepackage{xcolor}
\usepackage{soul}
\usepackage{graphicx}
\usepackage{dcolumn}
\usepackage{bm}
\newcommand\blfootnote[1]{%
	\begingroup
	\renewcommand\thefootnote{}\footnote{#1}%
	\addtocounter{footnote}{-1}%
	\endgroup
}
\graphicspath{{./Figures/}}

\begin{document}

\title{Traffic flow with multiple quenched disorders}

\author{A. Sai Venkata Ramana}
\affiliation{New York University Abu Dhabi, Saadiyat Island, P.O. Box 129188, Abu Dhabi, U.A.E.}

\author{Saif Eddin Jabari}
\email{Corresponding author email: \url{sej7@nyu.edu} \\}
\affiliation{New York University Abu Dhabi, Saadiyat Island, P.O. Box 129188, Abu Dhabi, U.A.E.}
\affiliation{New York University Tandon School of Engineering, Brooklyn NY.}


\begin{abstract}
We study heterogeneous traffic dynamics by introducing quenched disorders in all the parameters of Newell's car-following model.  Specifically, we consider randomness in the free-flow speed, the jam density, and the backward wave speed. The quenched disorders are modeled using beta distributions. It is observed that, at low densities, the average platoon size and the average speed of vehicles evolve as power-laws in time as derived by Ben-Naim, Krapivsky, and Redner (BKR). No power-law behavior has been observed in the time evolution of the second moment of density and density distribution function indicating no equivalence between the present system and the sticky gas. As opposed to a totally asymmetric simple exclusion process (TASEP), we found no power-law behavior in the stationary gap distribution and the transition from the platoon forming phase to the laminar phase coincides with the free-flow to congestion transition and is always of first-order, independent of the quenched disorder in the free-flow speed. Using mean-field theory, we derived the gap distribution of vehicles and showed that the phase transition is always of first-order, independent of the quenched disorder in the free-flow speed corroborating the simulation results. We also showed that the transition density is the reciprocal of the average gap of vehicles in the platoon in the thermodynamic limit.
\end{abstract}

\maketitle

\section{Introduction}
Traffic flow \blfootnote{NOTICE: this is the author’s version of a work that was accepted for publication in Physical Review E. Changes resulting from the publishing process, such as peer review, editing, corrections, structural formatting, and other quality control mechanisms may not be reflected in this document. Changes may have been made to this work since it was submitted for publication. A definitive version was subsequently published in Physical Review E DOI: \href{https://doi.org/10.1103/PhysRevE.00.002100}{10.1103/PhysRevE.00.002100}.} is a classic problem in non-equilibrium statistical physics that has attracted the attention of physicists and engineers for decades. Various phenomena pertaining to traffic have been studied using a variety of models~\cite{DC2000,Helbing2001,Nagel}. However,  the phenomena that the system exhibits are so complex that there are many aspects of it which are not yet completely understood. One interesting aspect which drew the attention of  physicists is the  emergence of collective behavior in the traffic system in the presence of a quenched disorder in a physical parameter which led to power-laws and phase transitions.

Ben-Naim, Krapivsky and Redner (BKR)~\cite{Ben1994} analytically derived the power-law behavior of various quantities like average speed, average platoon size (or length) of vehicles behind the slowest vehicle in the traffic stream when there is a quenched disorder in the free-flow speeds of the vehicles, which are the speeds of vehicles when unobstructed by other vehicles (also sometimes referred to as maximum speeds and desired speeds). They assumed that the vehicles (or the clusters of vehicles) move ballistically before interacting with their leader vehicle (cluster) and instantaneously merging with it. The assumed process is similar to the well-studied ballistic aggregation except that the momentum is not conserved in the present case. The system evolves into a stationary state in which there is a highly dense platoon of vehicles all moving at the same speed and a large gap ahead of the leader. BKR showed that power-laws for time evolution of the average platoon size behind the slowest vehicle and the average speed of the system have exponents which depend on the exponent of the free-flow speed distribution i.e., if $p(v) \sim v^{\mu}$ as $v \sim 0$, 
\begin{equation}
	N_{\mathrm{c}} \sim t^{\frac{\mu + 1}{\mu + 2}}, \label{lc-law}
\end{equation}
where $N_{\mathrm{c}}$ is the cluster size and 
\begin{equation}
    \langle  v \rangle \sim t^{-\frac{1}{\mu + 1}}. \label{v-law}
\end{equation}
Ben-Naim and Krapivsky went on further to develop a kinetic model for traffic flow~\cite{Ben1998,Ben1999} which also includes vehicle overtaking. 

Traffic flow has been extensively studied as a totally asymmetric simple exclusion process (TASEP) as originally introduced by Spitzer~\cite{Spitzer1970} and its variants.  Krug and Ferrari~\cite{Krug1996,Krug2000} studied the TASEP with quenched disorder in particle hopping rates and observed the formation of clusters with their size increasing with time. They conjectured that the power law for the cluster-size and speed should be similar to (Eq.~\eqref{lc-law} and Eq.~\eqref{v-law}).  Simulation studies by Bengrine et al.~\cite{bengrine1999} supported their conjecture.  A further confirmation came from the studies of Ktitarev et al.~\cite{Ktitarev1997} using a cellular automaton (CA) that is closely related to the TASEP with parallel updating of particle positions. Evans~\cite{Evans1996} independently studied the TASEP with particle-wise quenched disorder in the hopping rates and obtained exact solutions for the steady state of the system. Evans showed that the clustering of the vehicles is analogous to Bose-Einstein condensation.  Krug and Ferrari~\cite{Krug1996} also studied the phase transition from the low density jammed phase, where the platoon formation happens, to the high density laminar phase with no such platoon formation. They observed that the type of phase transition depends upon the distribution of the quenched disorder. In both works it is assumed that the quenched disorder in the hopping rates has the form
\begin{equation}
    f(x) \sim (x-c)^{\mu}. \label{hopp} 
\end{equation}
The transition was observed to be of first order for $\mu > 1$ and second order for $0< \mu \le 1$. It was also observed that this phase transition need not match the phase transition from the free-flow phase to the congested phase.

Car-following models are another popular way of simulating the traffic flow. While CA models of traffic are inspired by that of a lattice gas, car-following models are continuum models in space and time. A variety of car following models exist in the literature~\cite{Newell2002,Helbing2001,Treiber2013}.  These models, albeit time consuming to simulate, mimic real life traffic more  closely than the TASEP based models and can have physically meaningful model parameters. This advantage made the car-following models highly popular in engineering applications and there are commercial software packages which can simulate various complex traffic flow conditions using the car-following models~\cite{Vissim,Sumo}.

There are limited studies on heterogeneity in car-following and there exist studies on the effect of slow vehicles~\cite{masukura2007theory,Masukura_2009,li2016analytical}.  However, to our best of knowledge there exist no studies on the effects of quenched disorders on probabilistic car-following. Further, the effect of having quenched disorders in multiple parameters is not studied even in the TASEP and CA models, and remains unknown. Viewing the quenched disorder as a heterogeneity in the vehicles, the actual traffic system can have heterogeneity in various parameters in addition to free-flow speed.
For example, there can be heterogeneity in the minimum gap chosen by each driver, there can be heterogeneity in the reaction time of each driver and so on. The behavior of the traffic system would depend on the combined effect of various such parameters.  Thus, in this work, we study the power-law behavior in various dynamical quantities and the phase transitions in the Newell's car following model~\cite{Newell2002} by introducing quenched disorder into its parameters.

Newell's car-following model, quenched disorders and the simulation details are briefly discussed in Sec.~\ref{sec:Newell}. In Sec.~\ref{sec:powerlaws}, the power-laws in various quantities observed in the simulations of the Newell's model are discussed and analyzed.  In Sec.~\ref{sec:gapDistribution}, we study the stationary state gap distribution determined from the simulations and also develop a mean field theory to calculate it. In Sec.~\ref{sec:transition}, we study the phase transitions in the system using the simulations and analyze the results. The paper is concluded in Sec.~\ref{sec:conc}.

\section{The Newell's car following model and simulation details} \label{sec:Newell}
Car-following models generally involve conjectured relationships between a vehicle's speed (or equivalently their acceleration) and and the spatial gap between them and their leader and (in some models) the speeds of their leaders. Newell's car-following model conjectures that the speed of vehicle $i$ at time $t$ is a function of the spatial gap from their leader $i-1$ at time $t-\Delta t_{\mathrm{adapt}}$, where $\Delta t_{\mathrm{adapt}}$ is a speed adaptation time (or a reaction time).  The equation of motion is
\begin{equation}
	\frac{\mathrm{d} x_i(t)}{\mathrm{d}t} = v\big(s_i(t-\Delta t_{\mathrm{adapt}}) \big), \label{eqOfMot}
\end{equation}
where $s_i(t) \equiv x_{i-1}(t) - x_i(t)$ is the gap between vehicle $i$ and their leader $i-1$ at time $t$ and $v(\cdot)$ is a conjectured speed-spacing relation.  Newell's model basically assumes the following: (i) The path taken by the follower is a spatio-temporal translation of the path taken by the leader. (ii) The speed ($v$) of a vehicle depends on the gap ($s$) ahead of it as given below
\begin{equation}
v(s) = \begin{cases} v_{\mathrm{f}} & s > S_{\mathrm{c}} \\  \dfrac{ws}{S_{\mathrm{j}}} - w & S_{\mathrm{j}} \le s \le S_{\mathrm{c}} \end{cases}.
\label{vs}
\end{equation}
We chose Newell's car following model for our study as the model has only three parameters all of which are physically meaningful;
the jam density ($\rho_{\mathrm{j}}$), the free-flow (maximum) speed of the vehicle ($v_{\mathrm{f}}$) and the backward wave speed ($w$). It is the simplest car-following model with interpretable parameters and it has been empirically validated by others~\cite{ahn2004verification,4914846,chiabaut2010heterogeneous}.

At a microscopic level, the density $\rho$ perceived by a  driver may be interpreted as the reciprocal of the gap $s$ ahead of his/her vehicle. Thus, $S_{\mathrm{c}} = \rho_{\mathrm{c}}^{-1}$ is the critical gap, 
\begin{equation}
\rho_{\mathrm{c}} = \frac{w}{w+v_{\mathrm{f}}}\rho_{\mathrm{j}} \label{kc}
\end{equation}
is the critical density, and $S_{\mathrm{j}} = \rho_{\mathrm{j}}^{-1}$ is the reciprocal of jam density and is the minimum gap chosen by the driver. Eqs.(\ref{vs},\ref{kc}) essentially mean that a triangular fundamental diagram is assumed for each driver. The equations allow for a simple interpretation that the vehicles do not interact as long as $s > S_{\mathrm{c}}$. The interactions are present only when $s \le S_{\mathrm{c}}$ and are totally asymmetric i.e., the interaction effects only the follower.

The inputs to the simulation are (i) a track/circuit of chosen length, (ii) boundary conditions for the  track, (iii) the initial spatial and speed distributions of the vehicles, and (iv) the appropriate choices of the driver-vehicle related parameters $\rho_{\mathrm{j}}$, $v_{\mathrm{f}}$, and $w$. If all the driver-vehicle units were identical, $\rho_{\mathrm{j}}$, $v_{\mathrm{f}}$ and $w$ would be the same for all the vehicles. We employ a forward Euler scheme to solve \eqref{eqOfMot}: 
\begin{equation}
	x_i(t) = x_i(t - \Delta t) + \Delta t v \big( s_i(t - \Delta t) \big).
\end{equation}
The time step length used to simulate the system, $\Delta t$, is chosen in such a way that the maximum displacement is less than the minimum of $S_{\mathrm{j}}$ so that no accidents happen. The position of each vehicle is simultaneously updated using the simple kinematic equations following the speed rule of Eq.~\eqref{vs}. For the present work, we used periodic boundary  conditions for the track. The initial spatial distribution of the vehicles is taken to be uniform and each vehicle is given its maximum  speed so that the system is initialized in free-flow. Heterogeneity is introduced via a quenched disorder in each parameter of the model~\cite{jabari2014}. Thus, each vehicle has a different set of parameters drawn from a chosen distribution. This means that each vehicle has a different fundamental diagram depending on its parameters. It has been shown in previous work by one of us~\cite{jabari2014,jabari2018stochastic,zheng2018stochastic}  that the quenched disorders in the parameters of the model can be accurately described by a distribution which is bounded from above and below. The traffic flow problem naturally appeals for a heterogeneity of that kind. For example, the free-flow (maximum) speed of vehicles is expected to vary between a minimum non-zero value and a maximum value. Similarly, the minimum gaps chosen by the drivers and the backward wave speed have strictly positive minimum values and finite upper bounds. In our simulations, we use the beta distribution for the parameters. It is a bounded domain distribution that takes as special cases other bounded domain distributions. It can be tuned to be symmetric, left, or right-skewed, i.e., it offers flexibility in simulation. While it is not the only distribution with these particular features, it is the most widely-known distribution with these features, and it is a conventional choice for traffic variables~\cite{haight1963mathematical,jabari2020sparse}. The general form of the (standard) beta distribution is
\begin{equation}
    p_{\beta}(x) = \frac{1}{B(n,l)}x^{n-1}(1-x)^{l-1}I_{[0,1]}(x), \label{beta}
\end{equation}
where $B(n,l)$ is the beta function with parameters $n$ and $l$, and $I_{\mathcal{X}}(x)$ is the indicator function, which returns 1 if $x \in \mathcal{X}$ and returns 0 otherwise.  Let $A$ be any of the three parameters $v_{\mathrm{f}}$, $\rho_{\mathrm{j}}$ or $w$ with minimum and maximum values denoted by $A^{\min}$ and $A^{\max}$. The distribution of $A$, a generalized beta distribution is
\begin{multline}
	p_{A}(x) = \frac{(A^{\max} - A^{\min})^{-n-l+2}}{B(n,l)} \\ \times (x - A^{\min})^{n-1}(A^{\max}-x)^{l-1}   I_{[A^{\min},A^{\max}]}(x).  \label{beta1}
\end{multline}
In this work, we do not consider a quenched disorder in the speed adaptation time, $\Delta t_{\mathrm{adapt}}$.  This has subtle impacts that we will consider in a future paper.  In this paper, the delays are treated deterministically via the time step parameter $\Delta t$.

\section{Power-laws and finite size effects } \label{sec:powerlaws}
\subsection{Average cluster size and average speed}
In this section we study Newell's model with quenched disorders in its parameters $v_{\mathrm{f}}$, $\rho_{\mathrm{j}}$, and $w$. The system is prepared in a spatially homogeneous free-flow state.  The $v_{\mathrm{f}}$ for each vehicle is drawn from a beta distribution (cf. Eq.~\eqref{beta1}) with support $(v_{\mathrm{f}}^{\min}= 90,v_{\mathrm{f}}^{\max} = 110)$ in units of km/h. The jam density $\rho_{\mathrm{j}}$ is drawn from a beta distribution with support $(\rho_{\mathrm{j}}^{\min}=110, \rho_{\mathrm{j}}^{\max} = 170)$ in units of vehicles/km-lane.  The backward-wave speed is also drawn from a beta distribution with support $(w^{\min}=10,w^{\max} =30)$ in units of km/h. These values are typical for a highway. It has to be noted that the phenomena studied in the present  work depend only on the shape of the quenched disorders, not the parameters. For the discussions below, we used the beta distributions of $v_{\mathrm{f}}$ and $\rho_{\mathrm{j}}$, with $n=2$ and $l=2$ (symmetric distribution), and for the beta distribution of $w$, we chose $n=2$ and $l=3$ (skewed). Hereafter, we denote an  average over the quenched disorder using angular brackets ($\langle \cdot \rangle$) and an average over the system with a bar on the parameter ($\bar \cdot$). The superscripts and subscripts, wherever used, hold their usual meaning.  Specifically, if $\xi$ is a quenched disorder with domain $\Omega$ and probability density function $\mu_{\xi}$, then for any function of $\xi$, $\phi(\xi)$, we define the first average as
\begin{equation}
	\langle \phi(\xi) \rangle \equiv \int_{\omega \in \Omega} \mathrm{d} \omega \mu_{\xi}(\omega) \phi(\omega).
\end{equation}
In our simulations below, we calculate these averages by drawing a large number ($M$) of independent samples from the distributions of the quenched disorders, e.g., $\xi_1, \xi_2, \hdots, \xi_M$ and average over them, i.e., we use them to define an empirical distribution
\begin{equation}
	\mu_{\xi}(\omega) \approx \frac{1}{M} \sum_{1 \le m \le M}  \int_{\eta} \mathrm{d}\eta \delta(\eta - \xi_m)\delta(\eta-\omega)
\end{equation}
to approximate the average
\begin{multline}
	\langle \phi(\xi) \rangle \approx \frac{1}{M} \sum_{1 \le m \le M} \int_{\omega \in \Omega} d\omega   \int_{\eta} \mathrm{d}\eta \delta(\eta - \xi_m)\delta(\eta-\omega) \phi(\omega) \\ 
	= \frac{1}{M} \sum_{1 \le m \le M} \phi(\xi_m),
\end{multline}
where $\delta(\cdot)$ is the Dirac delta function.  The second average is taken over space in the system.  For example, let $\zeta(x)$ be a state variable defined over position in space, then
\begin{equation}
	\overline{\zeta} \equiv \frac{1}{L} \int_{\eta} \mathrm{d} \eta \int_0^L  \mathrm{d} x ~\eta \delta\big(\zeta(x) - \eta \big).
\end{equation}
The latter average can be interpreted as one that assumes an empirical probability density of the form $p_{\zeta}(\eta) = L^{-1} \int \mathrm{d} x \delta(\zeta(x) - \eta)$ and then calculates the average over this probability, i.e., $\overline{\zeta} = \int_{\eta} \mathrm{d} \eta p_{\zeta}(\eta)$.

\begin{figure}
	\resizebox{0.5\textwidth}{!}{%
	\includegraphics{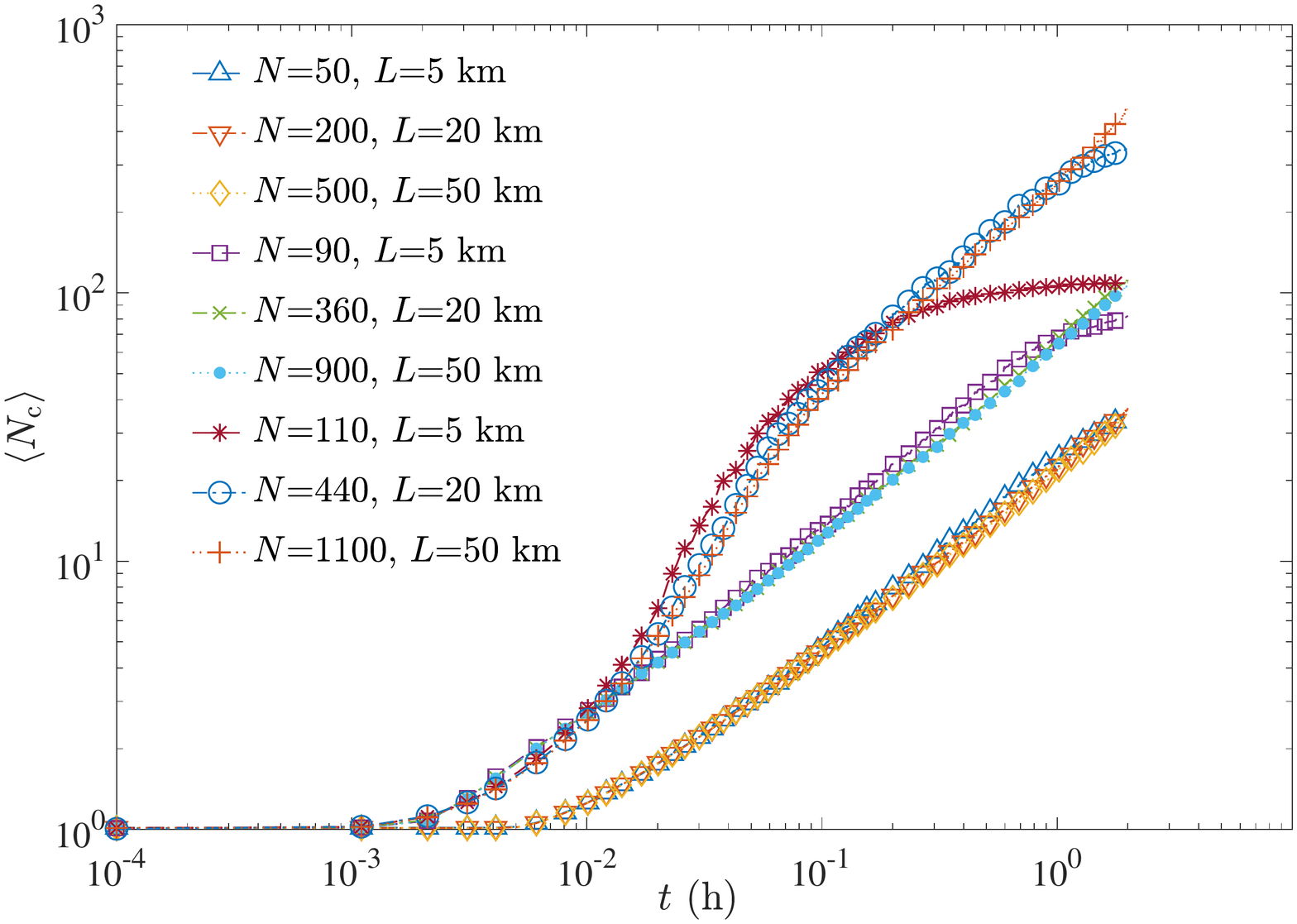}}
	\caption{\label{fig:avgcls} The average platoon size $\langle N_{\mathrm{c}} \rangle$ versus time $t$ for three densities on a log-log plot.  The platoon size is averaged over 1000 simulations to produce each of the curves in the figure, i.e., it is an ensemble average.  Platoon formation starts after a time $\tau_{\mathrm{f}}$. The linear curves for $t \gg \tau_{\mathrm{f}}$ imply a power law evolution for $\langle N_{\mathrm{c}} \rangle$.  We see plateauing in three of the curves towards the end. This is due to finite system size effects; all three cases correspond to experiments with smaller numbers of vehicles, $N = \{50,90,100\}$.  Outside of these finite population effects, we see a remarkable agreement between the curves in terms of rate growth of $\langle N_{\mathrm{b}} \rangle$ with time.}
\end{figure}

The average density $\bar \rho$ in the experiment below is chosen to be less than $\rho_{\mathrm{c}}^{\max}$, the maximum critical density from Eq.~\eqref{kc}, so that platooning occurs.  In Fig.~\ref{fig:avgcls} the average platoon size is plotted as a function of time for three different densities along three different circuits with varying lengths. The bottom two families of curves are for very low initial densities. 
Clearly, even for those cases, the initial state in which each vehicle moves at its own free-flow speed does not last long.
The average duration of the free-flow state $\tau_{\mathrm{f}}$ for each vehicle depends on the initial density $\overline{\rho} = NL^{-1}$, where $N$ is the number of vehicles, $L$ is the track length. The average duration free-flow $\tau_{\mathrm{f}}$ is defined as
\begin{equation}
	\tau_{\mathrm{f}} \equiv \left\langle \int_{t=0}^{\infty} \mathrm{d}t \Big[ 1 - H \Big( v_{\mathrm{f}} - v\big(s(t) \big) \Big) \Big] \right\rangle,
\end{equation}
where $H(\cdot)$ is the Heaviside step function.  In the thermodynamic limit ($N$ and $L$ get large but $\overline{\rho} = NL^{-1}$ is maintained), $\tau_{\mathrm{f}}$ and can be approximated as
\begin{equation}
	\tau_\mathrm{f} \approx \frac{\bar \rho^{-1} - \langle S_{\mathrm{c}} \rangle}{\langle v_{\mathrm{f}} \rangle - {v_{\mathrm{f}}}^{\min} }. \label{tau1}
\end{equation}
After time $\tau_{\mathrm{f}}$,  if $\bar \rho^{-1} \gg \langle S_{\mathrm{c}} \rangle$, the assumptions in the derivation of BKR hold for the present system and thus the system quickly enters into the power-law regime.  As $\bar \rho$ is increased, more vehicles start interacting with their leaders and a power-law describing the clustering process emerges once the speed adjustments resulting from these interactions become negligible. The uppermost family of curves in  Fig.~\ref{fig:avgcls} depict this.
 
\begin{figure}
	\resizebox{0.5\textwidth}{!}{%
	\includegraphics{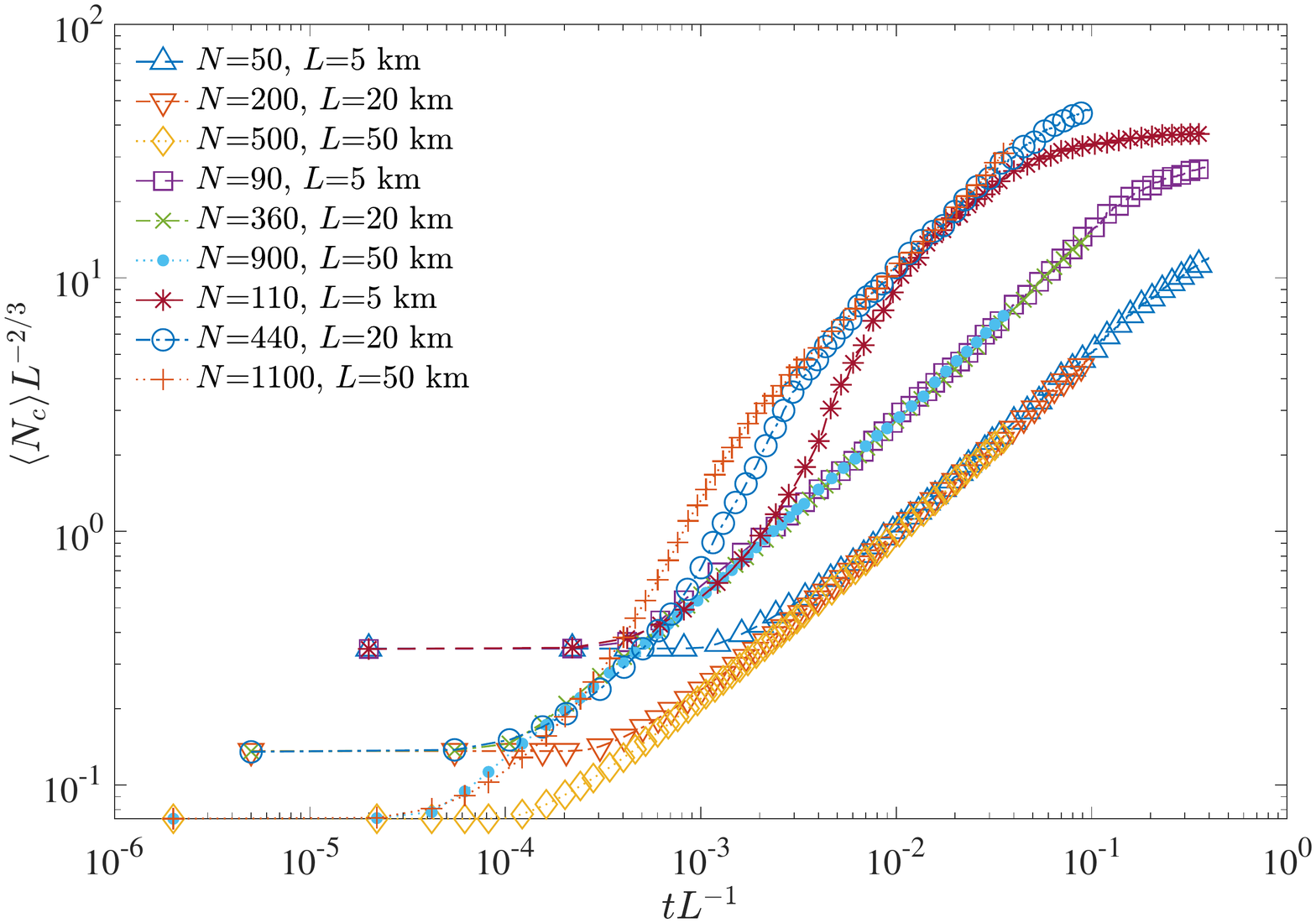}}
	\caption{\label{fig:scavgcls} Finite-size scaling of average platoon size (depiction same as Fig.~\ref{fig:avgcls}). The curves corresponding to same density collapse onto a single curve in the domain where the power-law holds, confirming the exponent to be $\alpha_{\mathrm{c}} = 2/3$ for the present case.}
\end{figure}

\begin{figure*}[th!]
\resizebox{0.76\textwidth}{!}{%
	\includegraphics{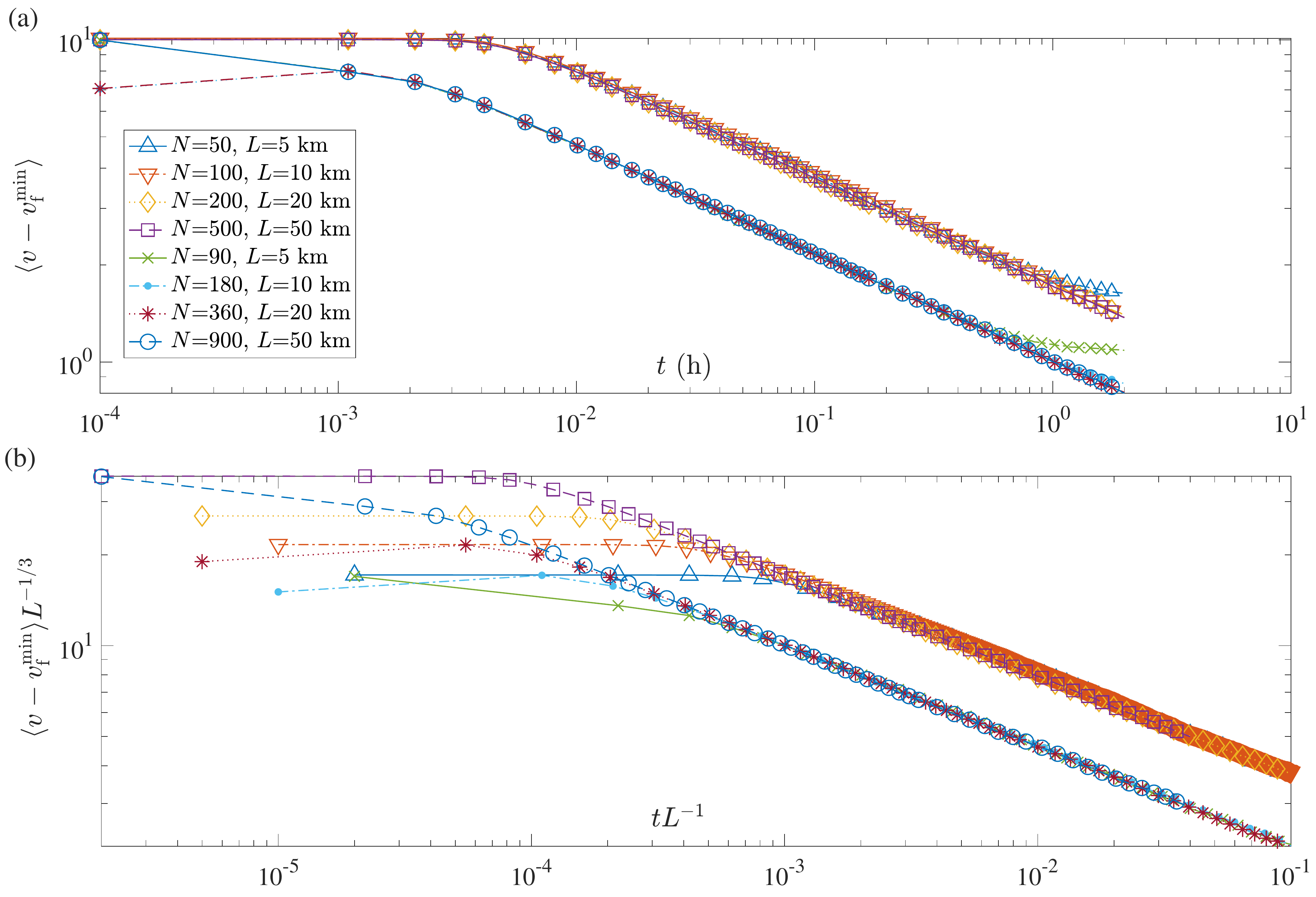}}
\caption{\label{fig:vav-nrt} Average relative speed versus time. (a): Average relative speed versus time for various traffic densities on a log-log plot. The linear curve for $t \gg \tau_{\mathrm{f}}$ implies a power-law. Deviation from linearity and flattening of the curves for the cases $N=50$ and $N=90$ are due to finite size effects. (b): Scaled average speed versus scaled time. Curves corresponding to the same density collapse onto a single curve in the domain where the power-law holds confirming the exponent to be $\alpha_{\mathrm{v}}=1/3$ for the present case.}
\end{figure*}

As the platoon size grows, deviations from the power-law are observed because of the finite number of vehicles and the finite track length (Fig.~\ref{fig:avgcls}). To account for the finite size effects, we assumed the following finite size scaling form for average platoon size:
\begin{equation}
	\langle N_{\mathrm{c}} \rangle = N^{\alpha_{\mathrm{c}}}f_{\mathrm{c}}(tN^{-1}), \label{fssl}
\end{equation}
where $f_{\mathrm{c}}(x)\sim x^{\alpha_{\mathrm{c}}}$. All the scaled curves collapse onto a single curve when $\alpha_{\mathrm{c}} = 0.67$ as can be seen from Fig.~\ref{fig:scavgcls}.
The value of $\alpha_{\mathrm{c}}$ exactly matches with that predicted by BKR. Similarly, we assume the finite size scaling form for relative speed
\begin{equation}
    \langle v - v_{\mathrm{f}} \rangle \sim N^{\alpha_v}f_{\mathrm{v}}(tN^{-1}),
\end{equation}
where $f_{\mathrm{v}}(x)\sim x^{\alpha_{\mathrm{v}}}$, and obtained $\alpha_{\mathrm{v}} = -0.33$ which matches with the analytical result of BKR. The collapse of the curves corresponding to same density to a single curve further confirms the scaling law (see Fig.~\ref{fig:vav-nrt}). 

\subsection{Relation to the sticky gas}
Sticky gas is a system of purely inelastic particles that stick and move together after collision. In one dimension, the problem has been exactly solved by mapping it onto the problem of shocks in the velocity field described by the inviscid Burger's equation~\cite{Frachebourg1999}. Thus, the power-laws corresponding to the sticky gas system are known exactly.  When the parameters of the beta distribution for $v_{\mathrm{f}}$ are $n=2$ and $l=2$, i.e., a symmetric beta distribution, the exponents of the power-laws of the average queue size and the speed in the present problem match those of the sticky gas and the granular gas~\cite{Shinde2009}. It is interesting to see if there is any equivalence in the universality classes between traffic flow and the sticky gas system for this case. Equivalence of both may imply a broader universality class to which traffic flow systems belong. 
 
Equivalence of both the systems should imply that the power-laws corresponding to all the quantities should be same for both the systems.  To this end, we investigate the behavior of the second moment and spatial correlation of local traffic densities, where the local density at $x$ can be interpreted as the reciprocal of gap between vehicles at position $x$. The probability density associated with observing a local traffic density of $\rho$ at any position along the road can be written as an empirical probability:
\begin{equation}
	p_{\varrho}(\rho,t) = \frac{1}{L} \int_0^L \mathrm{d} x~ \delta\big(\rho(x,t) - \rho\big)
\end{equation}
for which we employ a discrete space approximation to perform calculations:  We divide space into $N_{\mathrm{b}}$ blocks of equal size ($l_{\mathrm{b}}$).  Then the local density at time $t$ is $\rho(x,t) \approx \rho_{b(x)}(t)$, where $b(x)$ is the block containing position $x$. Then
\begin{equation}
	p_{\varrho}(\rho,t) = \frac{1}{N_{\mathrm{b}}} \sum_b \delta_{\rho_b(t),\rho}.
\end{equation}
The first and second moments of traffic density at time $t$ are
\begin{equation}
 	\langle \bar \varrho \rangle(t) = \left\langle \int \mathrm{d} \rho ~ \rho p_{\varrho}(\rho,t)\right\rangle = \left\langle \frac{1}{N_{\mathrm{b}}} \sum_b \rho_b(t) \right\rangle = \bar \rho
\end{equation}
and
\begin{equation}
\langle \bar \varrho^2 \rangle(t) = \left\langle \int \mathrm{d} \rho ~ \rho^2 p_{\varrho}(\rho,t)\right\rangle = \left\langle\frac{1}{N_{\mathrm{b}}} \sum_b \rho_b(t)^2\right\rangle.
\end{equation}
The spatial density auto-correlation function is defined as
\begin{equation}
    C_{\varrho\varrho}(\Delta x,t) = \frac{\langle \rho_b\rho_{b+\Delta x} \rangle (t)}{\langle \rho_b\rangle (t) \langle\rho_{b+\Delta x} \rangle (t)}.
\end{equation}

\begin{figure}[h!]
	\resizebox{0.5\textwidth}{!}{%
    \includegraphics{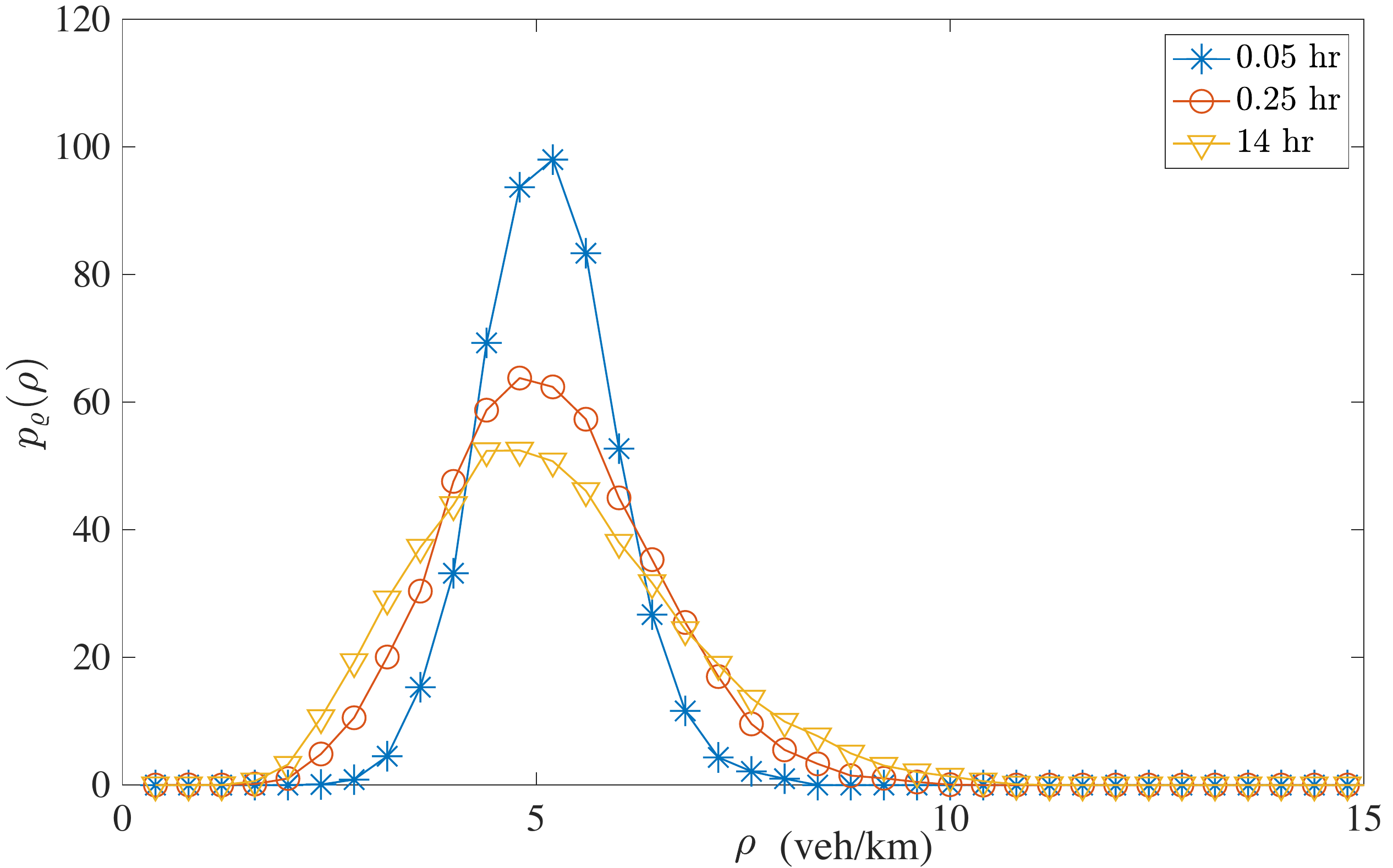}}
    \caption{ $p_{\varrho}(\rho,t)$ versus $\rho$ at various times ($N = 250$, $L = 5$ km, $l_{\mathrm{b}} = 100$ m, and $N_{\mathrm{b}} = l_{\mathrm{b}}^{-1}L = 50$) . The width of the distribution initially
    broadens but gets saturated to a constant value after a long time as the stationary state is reached. }
    \label{fig:PrhoVsrho}
\end{figure}

\begin{figure}[th!]
 	\resizebox{0.5\textwidth}{!}{%
 		\includegraphics{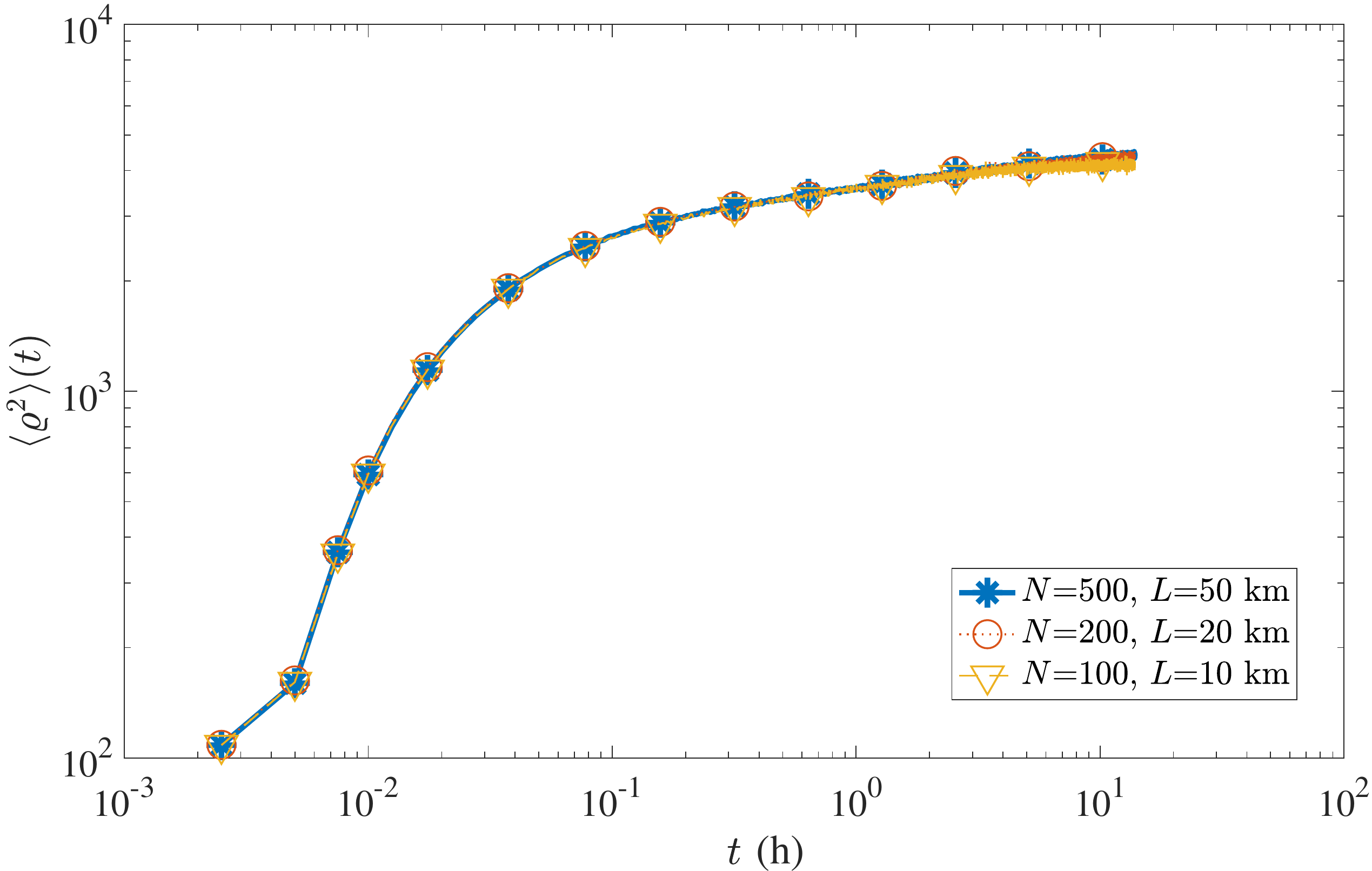}}
 	\caption{\label{fig:densqrt} Evolution of second moment of density. It can be easily seen that there is no power-law evolution.}
 \end{figure}

The $p_{\mathrm{sg}}(\rho,t)$ for a sticky gas, in the limit of large system size, scales with time as~\cite{Frachebourg1999}
 \begin{equation}
     p_{\mathrm{sg}}(\rho,t) = \frac{1}{t^{4/3}}f\left(\frac{\rho}{t^{2/3}}\right),
 \end{equation}
 where $f(z) \sim z^{-1/2}$ for $z \ll 1$ and $f(z) \rightarrow 0$ for $z \gg 1$. We find that the  $p_{\varrho}(\rho,t)$ determined from the simulations doesn't follow any power-law.  The calculated $p_{\varrho}(\rho,t)$ is shown in Fig.~\ref{fig:PrhoVsrho} for a particular case. It can be seen that the $p_{\varrho}(\rho,t)$ peaks around the average density. The width of the distribution increases with time and gets saturated as the stationary state is approached.  The second moment of density scales with time as $\langle \varrho^2 \rangle(t) \sim t^{2/3}$ for a sticky gas. The same is true for a granular gas in a limited regime~\cite{Shinde2009}. A typical $\langle \varrho^2 \rangle(t)$ from our simulation is shown in Fig.~\ref{fig:densqrt}. Clearly, $\langle \varrho^2 \rangle(t)$ doesn't follow any power-law which is again a distinction from expected behavior.  Further, in the case of a sticky gas~\cite{Bray2002}, the spatial auto-correlation is proportional to $t^{-2/3}$.  Interestingly, we find that this scaling law is obeyed in the present system. 
 \begin{figure}[th!]
 	\resizebox{0.5\textwidth}{!}{%
 		\includegraphics{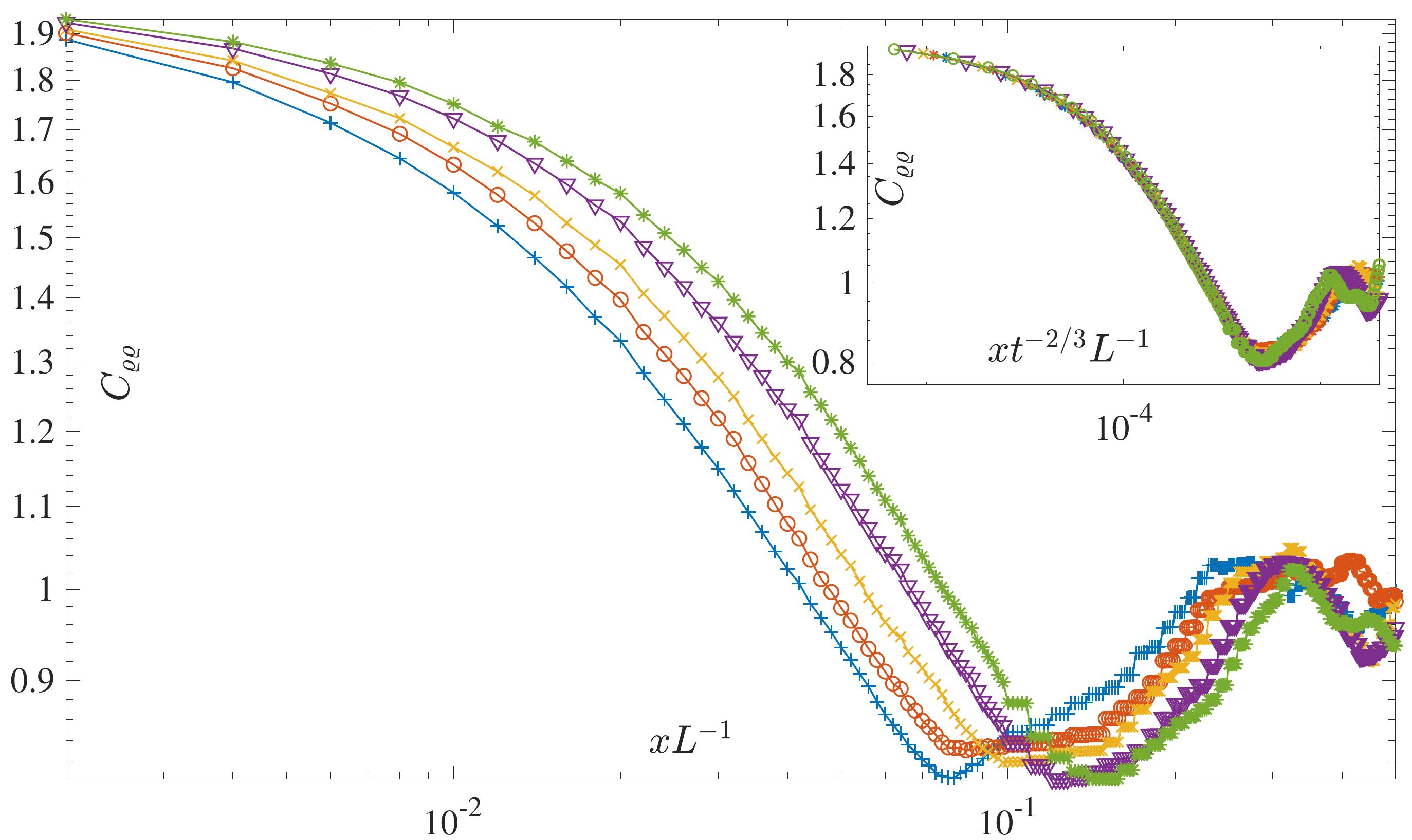}}
 	\caption{\label{fig:cddrt} Traffic density spatial auto-correlation function. The curves from bottom to top are in increasing order of time steps: $t$ = 1000, 2000 3000, 4000, 5000 respectively in units of $\Delta t \sim 10^{-5}$ hr. Inset: Scaled auto-correlation function.}
 \end{figure}
 
Thus we see that some quantities like the average platoon size, average speed, and spatial auto-correlation follow a power-law behavior that is similar to that of a sticky gas, while other quantities like the second moment of density do not. Hence there is no exact equivalence of the present system to a sticky gas and to the traffic flow models with just one quenched disorder. 
 
\section{Gap distribution} \label{sec:gapDistribution}
In this section, we determine from simulations the probability distribution of gap $p(s)$ for stationary states of the system in the regime where the BKR power-laws hold. We also develop a mean field theory to calculate $p(s)$.

\begin{figure}[h!]
\resizebox{0.5\textwidth}{!}{%
\includegraphics{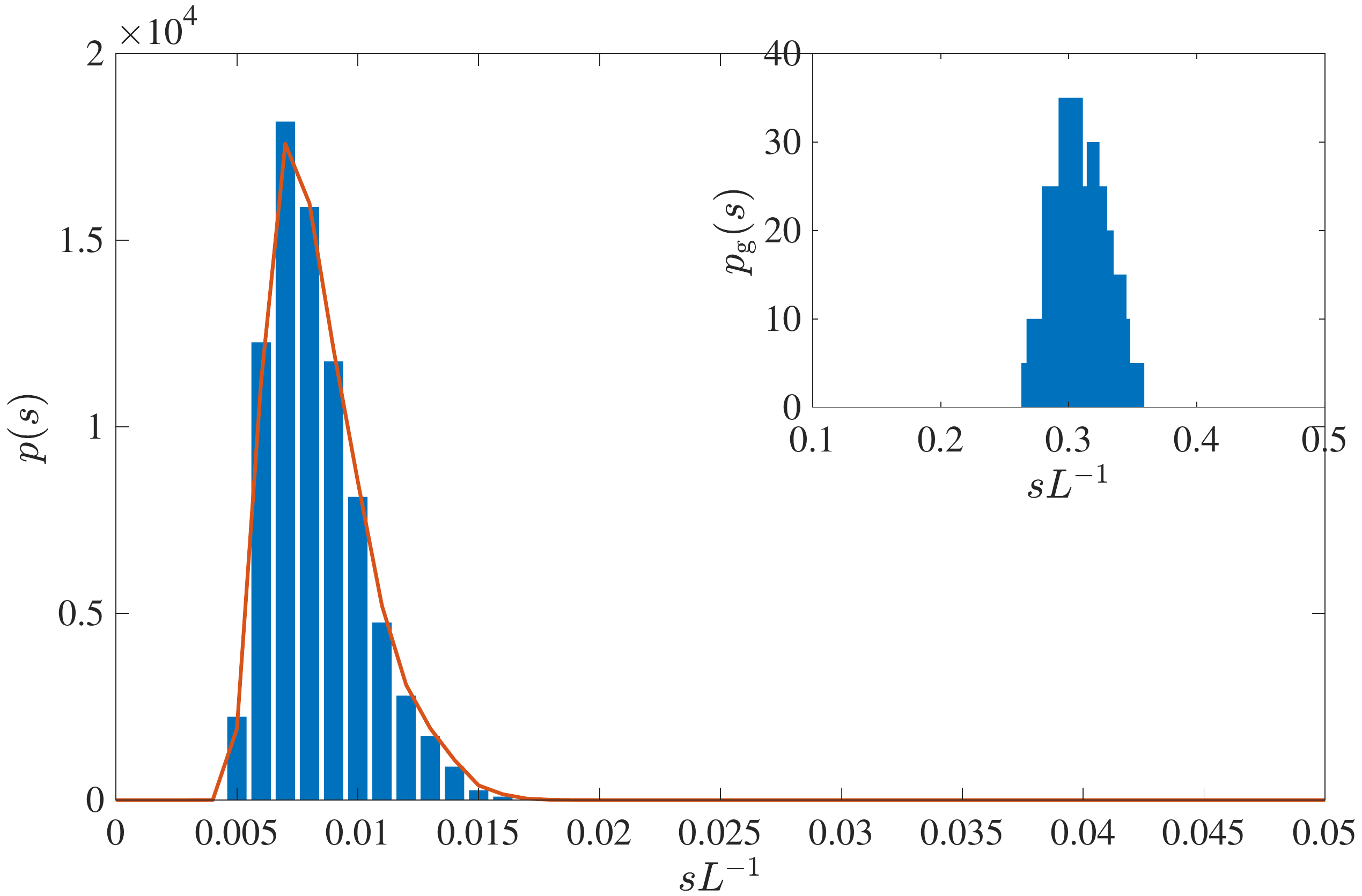}}
\caption{\label{fig:ps} Probability  of gap $p(s)$ for the stationary state of the system. The rectangular bins are from the simulation. The curve is $p_{\mathrm{c},l}(s)$. $p_\mathrm{g}(s)$ is shown in the inset. }
\end{figure}

To determine $p(s)$ from simulations, quenched disorder averaging is done over an ensemble of a few hundred copies of the stationary states of the system, depending on track length. A typical $p(s)$ we obtain is plotted in Fig.~\ref{fig:ps}. The distribution $p(s)$ has two distinct components that require separate treatments, the first component corresponds to gaps among non-leading vehicles ($p_{\mathrm{p}}(s)$), i.e., within the platoon; the other corresponds to the gap ahead of the slowest vehicle, which is greater than the maximum gap within the platoon ($S_{\mathrm{p}}^{\max}$). The two components, $p_{\mathrm{p}}(s)$ and $p_{\mathrm{g}}(s)$ are not probabilities but are supported on disjoint sets.  We may hence write
\begin{equation}
    p(s) = p_{\mathrm{p}}(s) + p_{\mathrm{g}}(s).
\end{equation}

Consider a (following) vehicle ($f$) moving at a speed $v = v_{\mathrm{f}} > v_{\mathrm{f},l}$ approaching a platoon, where $v_{\mathrm{f},l}$ is the free-flow speed of the leader ($l$) of the platoon. The follower $f$ starts interacting with the last vehicle in the platoon in accordance with its gap choice characteristics and thus starts slowing down so as to equalize its speed to that of $l$. From Eq.~\eqref{vs}, assuming $s>S_{\mathrm{j}}$, the gap from their leader upon joining the platoon (i.e., assuming the speed of the platoon leader) is
\begin{equation}
	s = \frac{v_{\mathrm{f},l} + w}{w} S_{\mathrm{j}}. \label{d3}
\end{equation}
This gap is equal to a critical gap of the platoon leader $l$: $s = S_{\mathrm{c},l}$. Thus, each vehicle in the platoon would
have a gap that is randomly chosen from the critical gap distribution of the leader of the platoon. We denote this gap by  distribution $p_{\mathrm{c},l}(s)$. The above arguments are true even if the distributions of the quenched disorder are correlated. However, the present work assumes the beta distributions of the $w$ and the $S_{\mathrm{j}}$ are independent and hence $p_{\mathrm{c},l}(s)$ is just the product of the beta distributions of $w$ and $S_{\mathrm{j}}$ given a $v_{\mathrm{f},l}$.

We now derive the form of the stationary state gap distribution applying mean field assumptions.  As the number of vehicles $N$ gets large, two things happen, all platoons  coalesce into a single platoon and the speed of the leader tends to the minimum free-flow speed, i.e., $v_{\mathrm{f},l} \rightarrow v_\mathrm{f}^{\min}$ in the stationary limit and as $N$ gets large. Consequently, in this regime, the speed distribution tends to a point mass at $p(v) \rightarrow \delta(v-v_\mathrm{f}^{\min})$. Thus the distribution of gap is
\begin{multline}
    p(s) = \int \mathrm{d}v ~ p(s|v)p(v) = p(s|v=v_\mathrm{f}^{\min}) \\= p_{\mathrm{p}}(s|v=v_\mathrm{f}^{\min}) + p_{\mathrm{g}}(s|v=v_\mathrm{f}^{\min})\\= p_{\mathrm{p}}(s) + p_{\mathrm{g}}(s).
\end{multline}
We dropped the dependence on $v=v_\mathrm{f}^{\min}$ in the last equality to simplify notation and as this should be understood from context.  For any vehicle selected at random from the $N$ vehicles in the system (i.e., chosen with probability $N^{-1}$), there are $N-1$ ways that a follower is chosen and one way that a leader is chosen.  Let $p_{\mathrm{p}}(s_1,\hdots,s_{N-1})$ denote the joint probability of the $N-1$ gaps in the platoon and $p_l(s)$ denote the probability density associated with the leader.  Then
\begin{multline}
p(s) = \frac{N-1}{N} \int \cdots \int \mathrm{d}s_2 \cdots \mathrm{d}s_{N-1} ~ p_{\mathrm{p}}(s,s_2\hdots,s_{N-1}) \\
+ \frac{1}{N}p_l \Big(s = L - \sum_{i=1}^{N-1}s_i \Big).
\end{multline}
Note that the indices in the above equation do not represent the order of the vehicles; under stationary conditions, the gaps between the vehicles in a platoon are governed by their individual quench disorders and are independent of one another. The first term in the sum is $p_{\mathrm{p}}(s)$ and the second term is $p_{\mathrm{g}}(s)$.  Since vehicles chose their gaps independently (under stationary conditions), the former simplifies to
\begin{multline}
	p_{\mathrm{p}}(s) = \frac{N-1}{N} \int \cdots \int \prod_{i=2}^{N-1}\mathrm{d}s_i ~ p_{\mathrm{c},i}(s_i) p_{\mathrm{c},l}(s) \\= \frac{N-1}{N} p_{\mathrm{c},l}(s) \int \cdots \int \prod_{i=2}^{N-1}\mathrm{d}s_i ~ p_{\mathrm{c},i}(s_i) = \frac{N-1}{N} p_{\mathrm{c},l}(s). \label{P_p}
\end{multline}
Hence, the gaps within the platoon are dependent entirely on the leader's critical gap choice.  That is, $S_{\mathrm{p}}^{\max} = S_{\mathrm{c},l}^{\max}$ (in the limit) and from Eq.~\eqref{kc}, 
\begin{equation}
    S_{\mathrm{c},l}^{\max} = \frac{v_{\mathrm{f}}^{\min} + w}{w}S_{\mathrm{j}}, \label{d4}
\end{equation}
which has a bounded domain as $S_{\mathrm{j}}$ and $w$ both have bounded domains. We next analyze the second component
\begin{equation}
p_{\mathrm{g}}(s) = \frac{1}{N}p_l \Big(L - \sum_{i=1}^{N-1} s_i \Big) \label{P_g1}
\end{equation}
by analyzing the limiting behavior as $N$ and $L$ get large but maintain the relation $NL^{-1}=\overline{\rho}$.  To do so, consider the distribution of
\begin{equation}
	S_l \equiv L - \sum_{i=1}^{N-1}S_{\mathrm{p},i},
\end{equation}
where $S_{\mathrm{p},i}$ is the $i$th gap within the platoon.  For large $N$ (and $L$), this approaches a normally distributed random variable with mean $\langle S_l \rangle = L - (N-1)\langle S_{\mathrm{p}} \rangle$ and variance $(N-1) \sigma_{\mathrm{p}}^2 \equiv (N-1) (\langle S_{\mathrm{p}}^2 \rangle - \langle S_{\mathrm{p}} \rangle^2)$.
Hence, for large $N$
\begin{multline}
	p_{\mathrm{g}}(s) \\ \approx \frac{1}{N} \frac{1}{\sqrt{2 \pi (N-1)\sigma_{\mathrm{p}}^2}} \exp \left( \frac{-(s - L + (N-1)\langle S_{\mathrm{p}} \rangle)^2}{2 (N-1)\sigma_{\mathrm{p}}^2} \right).
\end{multline}
That the variance of $S_l$ is proportional to $N$ implies that the variability in the gap ahead of the platoon leader grows as the circuit gets longer and the number of vehicles is increased. However, the variability in this gap in proportion to $L$ shrinks with $N$.  In other words, $S_l L^{-1}$ is normally distributed with mean $1 - (N-1)L^{-1}\langle S_{\mathrm{p}} \rangle \approx 1 - \overline{\rho}\langle S_{\mathrm{p}} \rangle$ and variance $(N-1)L^{-2}\sigma_{\mathrm{p}}^2 = (N-1)N^{-2}\overline{\rho} \sigma_{\mathrm{p}}^2$.  

It can be observed from the Fig.~\ref{fig:ps} that the stationary state distribution $p_{\mathrm{p}}(s)$ determined from our
simulations is similar to $p_{\mathrm{c},l}(s)$ as determined above and that $p_{\mathrm{g}}(s)$ resembles a normal distribution (with small variance for large $N$). The above arguments breakdown when the speed is no more $v_{\mathrm{f}}^{\min}$.
Thus, the average queue size and the average relative speed variations show power-law behavior as derived by BKR, which is an artifact of the quenched disorder in $v_{\mathrm{f}}$. The effects of quenched disorders in $\rho_{\mathrm{j}}$ and $w$ are that $p_{\mathrm{p}}(s)$ and $S_{\mathrm{c}}$ are bounded. We checked validity of the results for different values of the beta distribution parameters $n$ and $l$ corresponding to PDF of $v_{\mathrm{f}}$.

\section{Transition from power-law regime to non-power-law regime} \label{sec:transition}
As explained in the introduction, there are studies of phase transitions from the platoon formation phase to the laminar phase (where there is no platooning) in the TASEP and in the CA models. The TASEP simulations of Krug and Ferrari~\cite{Krug1996} and Bengrine et al.~\cite{bengrine1999} showed that the type of transition depends on the exponent of the distribution of hopping rates (Eq.~\eqref{hopp}). They also observed that the transition point for the phase transition from the platoon formation
phase to the laminar phase differs from the transition point for free-flow to congestion transition. Further, Krug and Ferrari showed that the gap distribution close to the transition is a power-law with the exponent depending on that of the hopping rate. Results of Ktitarev et al.~\cite{Ktitarev1997} corroborate those observations even though the quenched disorder is in deceleration rates.  In this section, we discuss the results of our simulations of the stationary states of the system and note the similarities and differences between our results and that of TASEP and CA models.

First we obtain the expressions for the average gap $\langle \bar S \rangle$ and the variance in the gap $\Delta_S^2$. We have that
\begin{multline}
	\langle S \rangle = \int \mathrm{d}s ~  s p(s)  = \int \mathrm{d}s ~ s \big(p_{\mathrm{p}}(s)+p_{\mathrm{g}}(s) \big) \\
	= \frac{N-1}{N}\langle S_{\mathrm{p}} \rangle + \frac{1}{N} \big( L - (N-1)\langle S_{\mathrm{p}} \rangle \big) = \frac{L}{N} \label{avgsg}
\end{multline}
as expected. Similarly,
\begin{multline}
	\Delta_S^2 = \int \mathrm{d}s ~(s - \langle S \rangle)^2 p(s) \\
	= \left(\int \mathrm{d}s ~ s^2 \big( p_{\mathrm{p}}(s) + p_{\mathrm{g}}(s) \big) \right) - \left(\frac{L}{N}\right)^2 
	= \frac{N-1}{N}\langle S_{\mathrm{c},l}^2 \rangle \\+ \frac{1}{N} \Big( (N-1)\langle S_{\mathrm{p}}^2 \rangle - \langle S_{\mathrm{p}} \rangle^2 + \big( L - (N-1)\langle S_{\mathrm{p}} \rangle \big)^2\Big) - \left(\frac{L}{N}\right)^2 \\
	= \frac{N-1}{N} \big( \langle S_{\mathrm{c},l}^2 \rangle + \langle S_{\mathrm{p}}^2 \rangle - \langle S_{\mathrm{p}} \rangle^2 \big) + \frac{1}{N} \langle S_{\mathrm{p}} \rangle^2 - \overline{\rho}^{-2} \\ + \frac{2}{\overline{\rho}} \big(1 - \overline{\rho} \langle S_{\mathrm{p}} \rangle \big)\langle S_{\mathrm{p}} \rangle + \frac{L}{\overline{\rho}} \big(1 - \overline{\rho} \langle S_{\mathrm{p}} \rangle \big)^2
\end{multline}
and for large $N$, we have that
\begin{multline}
	\Delta_S^2 \approx 2 \langle S_{\mathrm{c},l}^2 \rangle - \langle S_{\mathrm{c},l} \rangle^2 - \overline{\rho}^{-2} + \frac{2}{\overline{\rho}} \big(1 - \overline{\rho} \langle S_{\mathrm{c},l} \rangle \big)\langle S_{\mathrm{c},l} \rangle \\
	+ \frac{L}{\overline{\rho}} \big(1 - \overline{\rho} \langle S_{\mathrm{c},l} \rangle \big)^2, \label{vars}
\end{multline}
where we have used $\langle S_{\mathrm{p}} \rangle = \langle S_{\mathrm{c},l} \rangle$ in the limit based on the analysis above. 
From the last term, we see that the variance depends on the size of the system.  This is a result of the formation of 
a single platoon of vehicles behind the slowest vehicle and the mere existence of a gap in front of the platoon leader, which becomes unbounded in size as $L$ gets large.  As the density reaches the critical density of the platoon leader $\overline{\rho} \rightarrow S_{{\mathrm{c}},l}^{-1}$, the dependence on $L$ vanishes and $\Delta_S^2$ in this density regime is governed by intra-platoon gaps.  Indeed, as $\overline{\rho} \rightarrow S_{\mathrm{c},l}^{-1}$, the gap in front of the platoon leader disappears and the platoon leader ceases to be the platoon leader.  
\begin{figure*}[th!]
	\resizebox{0.76\textwidth}{!}{%
		\includegraphics{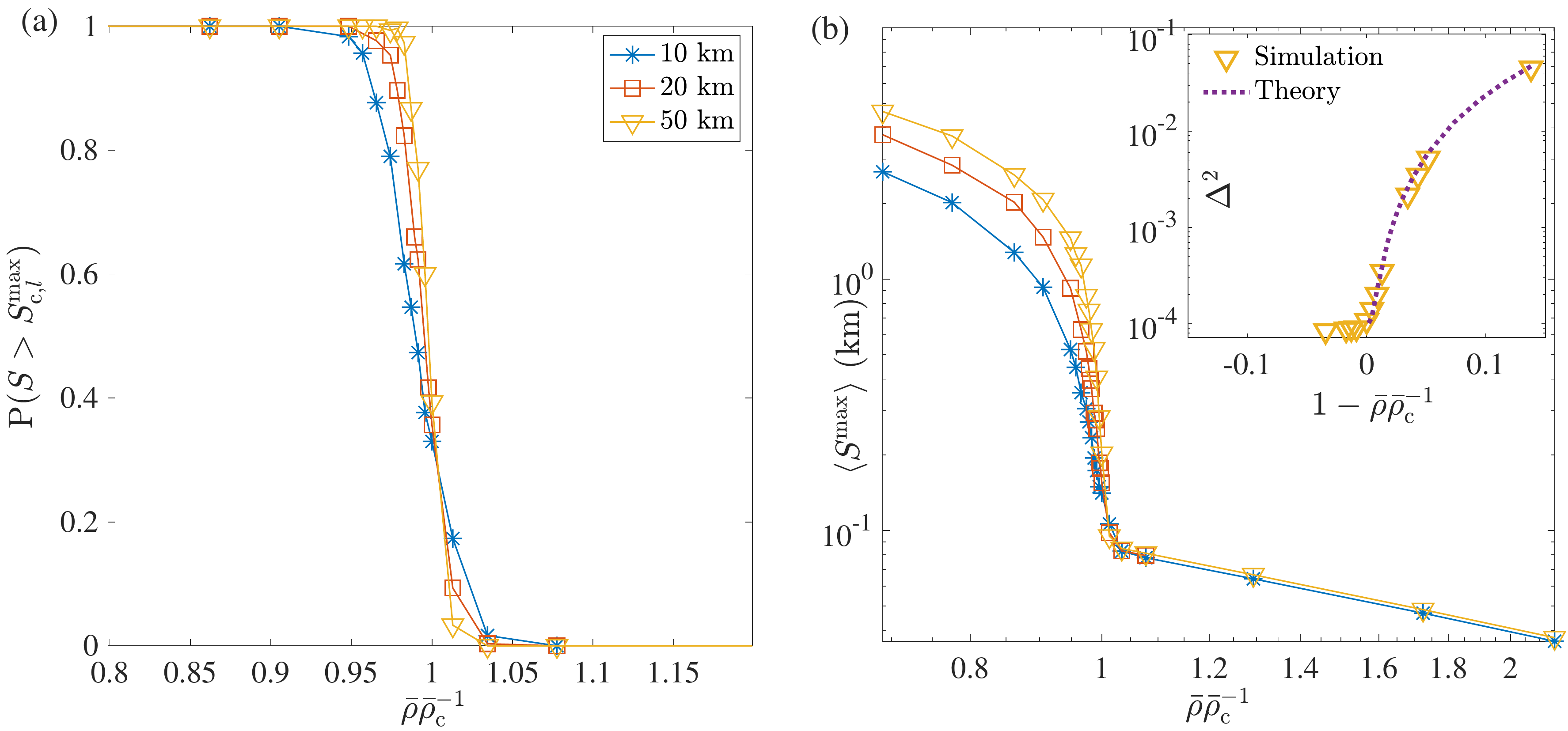}}
	\caption{\label{fig:avmaxgap} Maximum gap versus scaled average density.
		(a): $\mathbb{P}(S > S_{\mathrm{c},l}^{\max})$ versus $\overline{\rho}\overline{\rho}_{\mathrm{c}}^{-1}$ for various track lengths.
		(b): Ensemble average of maximum gap $\langle S^{\max} \rangle$ as a function
		of $\overline{\rho}\overline{\rho}_{\mathrm{c}}^{-1}$ for various track lengths. Inset: Gap variance $\Delta_S^2$ calculated from Eq.~\eqref{vars} (dotted line) along with one determined from simulation (triangles) for a track of length 50 km.
	}
\end{figure*}
Thus we take $\overline{\rho}_{\mathrm{c}} = \langle S_{\mathrm{c},l} \rangle^{-1}$ as the phase transition point and note that when $\overline{\rho} = \overline{\rho}_{\mathrm{c}}$ (i.e., at the phase transition), we have that 
 \begin{equation}
 	\Delta_S^2 \approx 2\big( \langle S_{\mathrm{c},l}^2 \rangle - \langle S_{\mathrm{c},l} \rangle^2 \big) \label{vars2}
 \end{equation}
in the thermodynamic limit ($N \rightarrow \infty$, $L \rightarrow \infty$ and $NL^{-1}= \overline{\rho}_{\mathrm{c}}$). Krug and Ferrari characterized the phase transition  to be of second order if the variance diverges as $\overline{\rho}_{\mathrm{c}}$ is approached.  If the variance is finite, the transition would be of first order. In the present case, the variance $\Delta_S^2$ becomes finite as the phase transition is approached as can be seen from Eqs.~\eqref{vars} and ~\eqref{vars2} implying that the transition is first order. As the above derivation doesn't use any information regarding the quenched disorder in $v_{\mathrm{f}}$ except for the fact that there exists a $v^{\mathrm{min}}_{\mathrm{f}}$,  the transition is expected to be of first order for any value of the parameters of the quenched disorder of $v_{\mathrm{f}}$.
 
Our simulations further confirm these results.  The simulations are done for various $\overline{\rho}$ close to $\overline{\rho}_{\mathrm{c}}$. In order to characterize the transition, we determined the following quantities as a function of $\overline{\rho}$: the average maximum gap $\langle S^{\max} \rangle$ in the system, the probability  of finding a gap greater than $S_{\mathrm{c},l}^{\max}$, i.e., $\mathbb{P}(S > S_{\mathrm{c},l}^{\max})$, $\Delta_S^2$, and the average flow $\langle J \rangle = \overline{\rho} \langle V \rangle$. 

\begin{figure*}[th!]
	\resizebox{0.76\textwidth}{!}{%
		\includegraphics{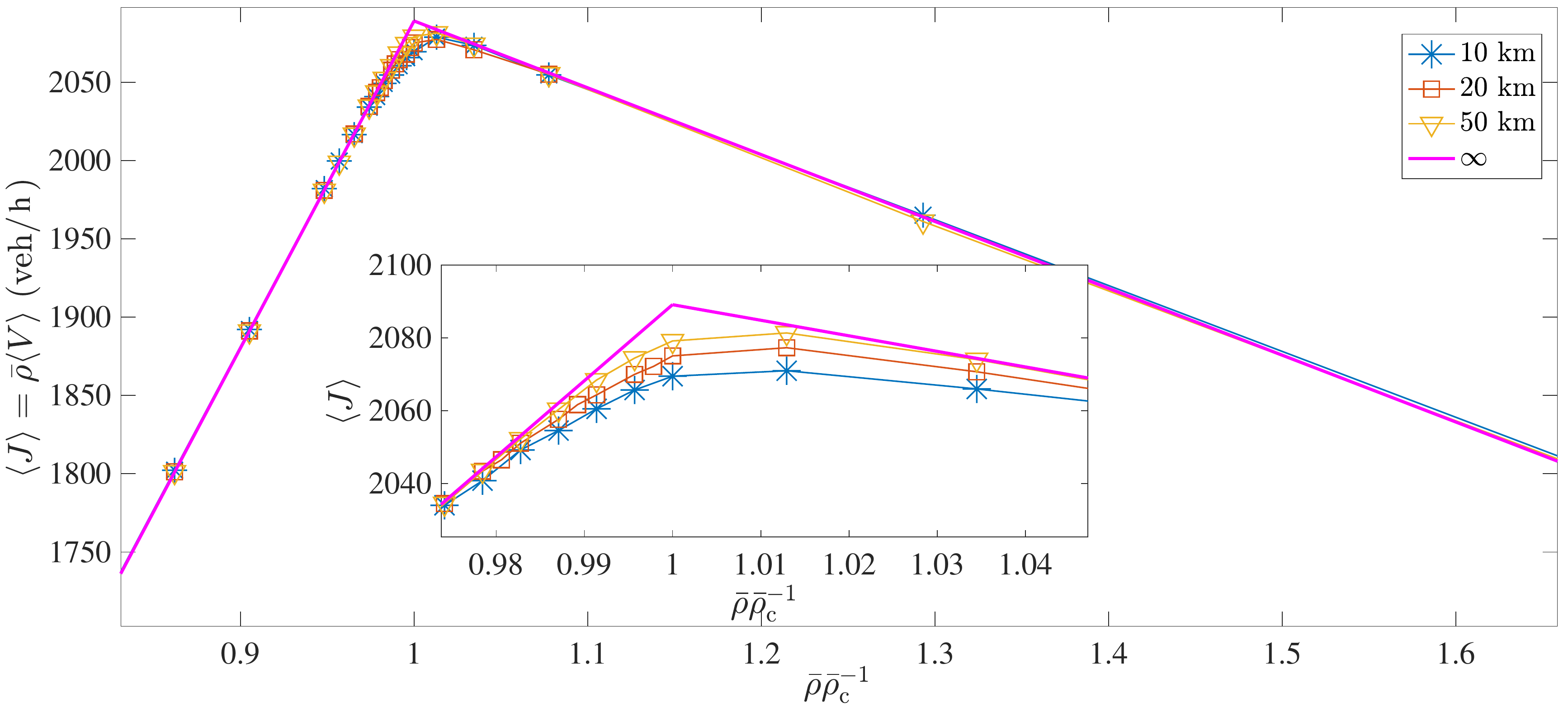}}
	\caption{Flow versus density diagram for various track lengths	when the parameters of PDF of speed are $n=2$ and $l=2$. The inset shows a magnified version of the diagram around the maximum flow.}
	\label{fig:jrho11}
\end{figure*}
The left pane of Fig.~\ref{fig:avmaxgap} shows $\mathbb{P}(S > S_{\mathrm{c},l}^{\max})$ for various track lengths.  The probability decreases sharply as the transition density $\overline{\rho}_{\mathrm{c}}$ is approached.  The ensemble averaged maximum gap  $\langle S^{\max} \rangle$ also decreases sharply as $\overline{\rho}_{\mathrm{c}}$ is approached. See the right pane of  Fig.~\ref{fig:avmaxgap}. The sharpness increases as the track length $L$ is increased indicating a discontinuity in the thermodynamic limit.  In the inset of Fig.~\ref{fig:avmaxgap}, the variances calculated from simulation and the theory (Eq.~\eqref{vars}) for a track of length 50 km are shown.  An excellent match between them solidifies our theoretical findings. Thus, all the quantities we determined classify the transition as a first order transition. The flow-density diagram is plotted in Fig.~\ref{fig:jrho11}.

\begin{figure*}[th!]
	\resizebox{0.76\textwidth}{!}{%
	\includegraphics{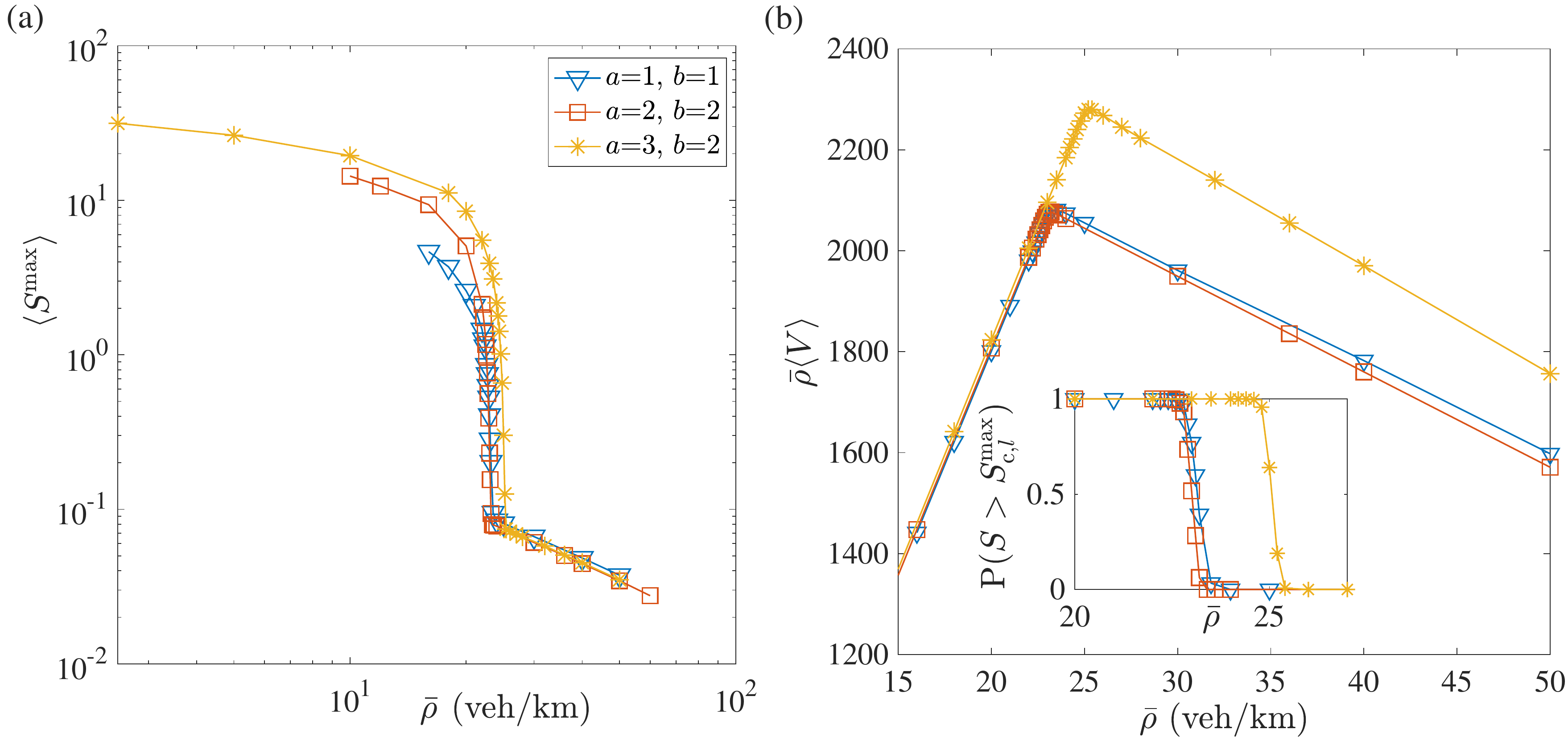}}
	\caption{Maximum average gap versus density for a track of length 50 km and three sets of parameters of the quenched disorder for $v_{\mathrm{f}}$. (a): $\langle S^{\max} \rangle$ versus $\overline{\rho}$. (b): $\langle J \rangle$ versus $\overline{\rho}$. Inset: $\mathbb{P}(S>S_c^{\max})$ versus $\overline{\rho}$.}
	\label{fig:ndep}
\end{figure*}

The sharpness of the free-flow to congestion transition also increased with track length. The transition point in the thermodynamic limit is obtained by linearly extrapolating the free-flow curve and the congestion curves away from the transition point and by determining their intersection point.  It can be seen that the transition point in the thermodynamic limit exactly matches with the predicted value of $\overline{\rho}_{\mathrm{c}}$.  Thus, both the phase transitions happen at the same traffic density. Eq.~\eqref{vars} suggests that it would be first order for any values of parameters in quenched disorder for $v_{\mathrm{f}}^{\min}$. We confirmed it by performing simulations with other values of the parameters. See Fig.~\ref{fig:ndep}.

All the above results are in sharp contrast to those observed in the TASEP~\cite{Krug1996,bengrine1999} and CA~\cite{Ktitarev1997} simulations.  Another interesting point to observe is that the triangular form of the flow-density diagram is preserved even after heterogeneity is included which says that the heterogeneous system may be mapped onto an effective homogeneous system with a suitable choice of parameters. 

\section{Summary and Discussion} \label{sec:conc}
We have simulated the traffic system in its full generality within Newell's model by incorporating quenched disorders in
 the free-flow speed, the jam density and the kinematic wave speed.  The quenched disorders are modeled using beta distributions.  We observed that system follows power-laws for average platoon size and the average speed as derived by Ben-Naim et al. We found that the gap distribution is a bounded domain distribution. To see if there is any equivalence between the sticky gas and the present system, we calculated  the second moment of local traffic density, the density distribution and the spatial auto-correlation in local traffic densities for a particular set of parameters. We observed that the second moment of density and the density distribution do not follow a power-law indicating no equivalence between the present system and the sticky gas. As opposed to TASEP, there was no power-law behavior in the gap distribution at the critical density corresponding to a phase transition from the platoon formation phase to the laminar phase.  Using mean field assumptions we derived the form of the gap distribution in the system in the platoon formation phase. It turns out that the gap distribution of vehicles in the platoon in the thermodynamic limit is simply the critical gap distribution of the leader. It is shown that the variance becomes finite as the phase transition point is approached and thus the transition is always of first order independent of the form of the quenched disorder, which is again in contrast to TASEP results. The transition density is shown to be nothing but the reciprocal of the average gap in the platoon. Simulation results support our theory.

Overall, the present results do not exactly conform with those of TASEP with quenched disorder. However, we expect that the present simulations closely mimic the real picture than the TASEP and related models. Thus the results of this work may be viewed as a benchmark for coarse grained models which use hydrodynamics or kinetic theory and incorporate heterogeneity.

\begin{acknowledgments}
This work was supported by the NYUAD Center for Interacting Urban Networks (CITIES), funded by Tamkeen under the NYUAD Research Institute Award CG001 and by the Swiss Re Institute under the Quantum Cities\textsuperscript{TM} initiative.
\end{acknowledgments}

\bibliographystyle{apsrev4-2}

\providecommand{\noopsort}[1]{}\providecommand{\singleletter}[1]{#1}%

\end{document}